\pdfoutput=1

\documentclass[fleqn,10pt]{article}
\usepackage[utf8]{inputenc}
\usepackage[T1]{fontenc}
\usepackage{graphicx}
\usepackage{siunitx}
\usepackage{authblk}
\usepackage[HTML]{xcolor}
\usepackage{soul}
\definecolor{colA}{HTML}{fdae61}
\definecolor{colB}{HTML}{f46d43}
\definecolor{colC}{HTML}{d53e4f}
\definecolor{colD}{HTML}{9e0142}
\definecolor{colE}{HTML}{238b45}
\definecolor{colF}{HTML}{3288bd}
\definecolor{colG}{HTML}{5e4fa2}
\usepackage{amsmath}
\usepackage{hyperref}
\usepackage{verbatim}
\usepackage[square,numbers,sort&compress]{natbib}
\usepackage{ulem}
\usepackage{setspace}
\usepackage{float}
\usepackage{booktabs}
\usepackage[margin=2.5cm]{geometry}
\newcommand{\mi}{\ensuremath{\mathbf{M}_\textsc{I}}}
\newcommand{\mii}{\ensuremath{\mathbf{M}_\textsc{II}}}
\newcommand{\cumd}{\ensuremath{\mathrm{C}\mu\mathrm{MD}}}

\DeclareSIUnit\Molar{\textsc{m}}
\DeclareSIUnit\elementarycharge{\text{\ensuremath{e}}}
\DeclareSIUnit\bar{bar}
\DeclareSIUnit\angstrom{\text{Å}}

\title{Ionic gradients shape the spatial organization and dynamics of supramolecular polymers}

\author[1,*]{Claudio Perego}
\author[2]{Annalisa Cardellini}
\author[1]{Andrea Tagliabue}
\author[1]{Giovanni M. Pavan}

\affil[1]{Department of Applied Science and Technology, Politecnico di Torino, Corso Duca degli Abruzzi 24, 10129 Torino, Italy}
\affil[2]{Department of Innovative Technologies, University of Applied Sciences and Arts of Southern 
Switzerland, Polo Universitario Lugano, Campus Est, Via la Santa 1, 6962 Lugano-Viganello, Switzerland}
\affil[*]{corresponding author: Claudio Perego (claudio.perego@polito.it)}
\date{}
\begin{document}
\doublespacing

\maketitle

\section*{Significance}

Living supramolecular structures such as actin filaments and microtubules assemble and function in spatially heterogeneous chemical environments shaped by concentration gradients. Engineering synthetic materials with comparable environmental coupling is a central challenge, yet gradient-driven supramolecular dynamics remains poorly understood. Here, we extend constant chemical potential molecular dynamics (\cumd) to simulate coarse-grained models of reversible supramolecular polymerization upon sustained ionic gradients. We consider two intrinsically different supramolecular polymer systems, revealing how ionic gradients generate spatially organized states, which exhibit local dynamics inaccessible in homogeneous systems with the same overall composition. More broadly, our approach enables the simulation of responsive supramolecular materials under realistic, chemically heterogeneous environments, opening a route to rational design and simulation of gradient-responsive materials.

\begin{abstract}

Living supramolecular materials such as microtubules and actin filaments assemble and function in dynamic, heterogeneous environments, subject to chemical concentration gradients. Their rapid and precise response to such gradients---via aggregation, disassembly, or autonomous motion---underlies the fine spatiotemporal control typical of living systems. Reproducing such responsiveness in synthetic materials is a central challenge in materials science and requires a molecular-level understanding of how gradients couple to supramolecular structure and dynamics, which remains difficult to access experimentally. To this end, we combine implicit-solvent coarse-grained molecular dynamics (MD) with the constant chemical potential MD (C$\mu$MD) technique to investigate, at molecular resolution, the coupling of two prototypical supramolecular polymer models with externally imposed ionic concentration gradients.
Despite their contrasting responses to homogeneous ionic conditions, both systems undergo pronounced spatial reorganization under the gradient, generating regions of locally enhanced supramolecular activity that are inaccessible to homogeneous systems with the same overall content of self-assembling species. Crucially, the local supramolecular dynamics is largely determined by the local concentrations, consistent with the formation of local quasi-equilibria: this direct connection between spatial composition and supramolecular activity emerges as a common governing principle across the two models. These findings shed light on the spatially complex supramolecular response that can be triggered by chemical gradients, and show how a single gradient simulation can encode this response across a continuous range of local chemical environments, thus providing a general strategy to investigate and design materials operating under spatially heterogeneous conditions.

\end{abstract}

\section*{Introduction}

Living materials operate out of equilibrium and often in spatially heterogeneous chemical environments, where molecular concentration gradients control their structure and dynamics~\cite{Harold2005,Bastiaens2006,Epstein2016,Das2021,Nguindjel2022}. Representative examples include the regulated assembly and disassembly of transient supramolecular structures such as microtubules and actin filaments~\cite{Kerssemakers2006,Pollard2009,Loose2014}, as well as the formation of membraneless organelles through liquid--liquid phase separation, which can be controlled by local variations in molecular concentration~\cite{Brangwynne2009,Sing2020}. Reproducing this spatially controlled responsiveness to the chemical environment in synthetic materials is a major goal of materials science.

Synthetic supramolecular polymers~\cite{Lehn1993,DeGreef2009,wehner2020supramolecular} provide a versatile platform for the design of materials that reproduce some of the dynamic, responsive, and functional features of living systems~\cite{Aida2012,merindol2017}. The reversible, noncovalent nature of their interactions enables properties such as stimuli-responsiveness, self-healing, and structural adaptability~\cite{Yan2012}, making them attractive for applications in medicine~\cite{Alvarez2021_spinalcord}, molecular transport and delivery~\cite{Qiu2022_supramolecular,Correa2022_liposome},
and catalysis~\cite{Neumann2015_organocatalysis,Cardona2020}.

Building on these dynamic properties, a broad range of biomimetic supramolecular systems has been developed in which transient assemblies are maintained out of equilibrium, including fibers~\cite{Mishra2018}, micelles~\cite{Maiti2016,Cardona2020,Pryanka2024_ATPgradient}, droplets~\cite{nakashima2018_coacervate,spath2021_peptidepolyelectrolyte}, and macroscopic gels~\cite{bal2019_fueleddissipative}. A key challenge is to understand quantitatively how the structure and dynamics of such supramolecular assemblies respond and couple to spatial chemical concentration gradients. Beyond biomimicry, this understanding is directly relevant to microfluidic experiments, where diffusion, mixing, and flow can be controlled to generate concentration gradients and steer self-assembly pathways toward non-equilibrium structures~\cite{sorrenti2016supramolecularchiral, rubiomartinez2016gradients, sevim2018microfluidic}.

Interest in chemically structured and dynamically varying environments has grown rapidly in recent years~\cite{Nguindjel2022,Velegol2016,Ashkenasy2017,Budroni2022,Araujo2023}, with an increasing number of supramolecular systems reported to exploit concentration gradients and fluctuations to generate behaviors such as autonomous motion, spatial patterning, and chemical signaling~\cite{Illien2017_fuelled,Chivers2019_shaping,vanderWeijden2020_autonomous,Maity2021_feedback}. 
Notable examples include the use of differential diffusion in gel matrices to spatially regulate the self-assembly of hydrazone molecular
gelators~\cite{Lovrak2019}, and ATP gradients that control micelle formation in metal--amphiphile systems~\cite{Cardona2020,Pryanka2024_ATPgradient}. These studies show that sustained chemical gradients can direct supramolecular assembly in space, generating structured non-equilibrium states that cannot be reproduced under homogeneous conditions at the same overall composition~\cite{Juneja2024_microfluidicgradient,Yadav2024_saltgradientinduced}. 

In parallel, studies of fueled dissipative supramolecular materials~\cite{Mishra2018,Cardona2020,sorrenti2017a,Ragazzon2018,Riess2020} have shown how temporal variations in chemical composition can regulate transient assembly. In these systems, the interplay between assembly, disassembly, and fuel consumption controls the lifetime and morphology of the resulting structures. Furthermore, the coupling between concentration gradients and supramolecular kinetics can give rise to collective dynamical phenomena, including self-assembly and degradation waves, oscillatory behaviors~\cite{Kubota2020,Leira2018}, and evolving fluorescent patterns~\cite{Cao2023}. Some of these artificial systems have been explicitly inspired by the spatiotemporally regulated assembly and disassembly dynamics of cytoskeletal filaments~\cite{Kubota2020}. More broadly, such spatially segregated  polymerization and depolymerization is reminiscent of treadmilling in active polar filaments, where growth and shrinkage occur at opposite ends of the same filament~\cite{Wegner1976,Erlenkaemper2013}.

The experiments discussed above typically probe supramolecular responses over micrometer-to-millimeter length scales, while the underlying molecular processes are not directly resolved and must be inferred indirectly. This limitation largely reflects the limited spatiotemporal resolution of available experimental techniques, although recent approaches based on super-resolution microscopy are beginning to provide \textit{in situ} access to information at molecular scale~\cite{Adelizzi2018_superresolution,
Dhiman2022_superresolution}. In this context, molecular simulations offer a complementary route to resolve the microscopic mechanisms underlying these collective phenomena.

Molecular dynamics (MD) simulations have become a key tool for elucidating the molecular mechanisms that govern supramolecular materials~\cite{bochicchio2018a}. They span a wide range of resolutions, from all-atom simulations, which describe supramolecular interactions and dynamics with high chemical specificity~\cite{Tantakitti2016,cardellini2025_exchange,Obenauer2024_atomistic}, to coarse-grained (CG) models, which trade atomistic resolution for access to larger length and time scales and therefore to slower collective processes~\cite{Lee2012,Bejagam2015,bochicchio2017from}. In particular, CG-MD has enabled extensive characterization of the structural and dynamical properties of biomimetic supramolecular polymers, including chemically fueled systems~\cite{Baker2015,Garzoni2016,Schoenmakers2018,bochicchio2017from,bochicchio2017into,bochicchio2017effect,demarco2021,Mishra2018,Kumar2014}. %

A further challenge in molecular simulations of supramolecular systems is the mismatch between the concentrations that can be efficiently accessed in finite simulation boxes and those commonly used experimentally. Even in CG models, simulated monomer concentrations are often substantially higher than experimental values: the models considered here operate at
$\sim~$20--80~mM, whereas experimental supramolecular systems are typically studied in the $\sim~$10~$\mu$M--1~mM range. Implicit-solvent
representations~\cite{Arnarez2015} help mitigate this limitation by reducing the number of degrees of freedom and extending the accessible time and length scales, thus allowing the simulation of much lower concentrations. They have been employed successfully to investigate supramolecular polymer dynamics~\cite{bochicchio2017into,demarco2021,Perego2021,crippa2022molecular}, colloidal self-assembly~\cite{Bianchi2019,Prestipino2019}, and amyloid aggregation~\cite{saric2014,Michaels2020}.

Despite the growing use of molecular simulations to investigate supramolecular systems, the response to spatial concentration gradients has so far been addressed primarily through continuum theoretical descriptions, using mass-balance
models~\cite{knowles2009,TenEikelder2019,Cohen2011} or reaction--diffusion frameworks~\cite{Epstein2016,Kubota2020,Leira2018,Cao2023}. These approaches successfully capture spatiotemporal phenomena at mesoscopic and macroscopic length scales  but necessarily describe the underlying molecular processes through effective kinetic and transport parameters. As a result, the molecular-level mechanisms by which supramolecular assemblies respond to chemical gradients remain largely unresolved.

In this work, we address this gap by investigating, at molecular resolution, the response of supramolecular polymer systems to externally imposed ionic concentration gradients. We employ implicit-solvent CG models~\cite{Perego2021,crippa2025nontrivial}, which provide access to the length and time scales relevant to supramolecular dynamics while retaining sufficient resolution to capture the processes governing assembly and disassembly. To impose and sustain the ionic concentration gradients, we use the constant chemical potential MD ($\cumd$) technique~\cite{Perego2015,Ozcan2017,Karmakar2023}, which maintains different local chemical environments across the simulation box and establishes a stationary, spatially resolved concentration profile. $\cumd$ has previously been applied to gas permeation through porous materials~\cite{Ozcan2017,ozcan2020_transport}, gas solubility in liquids~\cite{ansari2020_gassolubility}, crystal growth and morphology control~\cite{Perego2015,bjelobrk2019,karmakar2018a,Neha2024_crystalshape}, and covalent polymerization~\cite{Chen2020_polyandcryst}. Here, for the first time, we extend this framework to probe the response of supramolecular polymers to sustained concentration gradients.

We consider two prototypical supramolecular polymer models characterized by distinct modes of coupling to ionic species. Model~I describes a cooperative supramolecular polymerization driven by inter-monomer dipole--dipole interactions, in which
ions only weakly couple to the interactions responsible for assembly. Model~II instead describes a fuel-mediated supramolecular polymerization, in which positively charged monomers are bridged by multivalent fuel molecules, while dissolved ions compete directly with the fuel for monomer binding sites. The two models therefore provide contrasting examples of indirect and direct ionic coupling to supramolecular assembly.

Despite their markedly different responses to homogeneous ionic conditions, both systems develop pronounced spatial organization when exposed to a sustained ionic concentration gradient. Interestingly, a related gradient-induced spatial reorganization has recently been observed experimentally in supramolecular vesicle
systems~\cite{Yadav2024_saltgradientinduced}.
In Model~I, uniformly distributed ions have only
a weak influence on assembly, whereas the gradient induces a substantial redistribution of the monomers, creating locally enriched regions with structural and dynamical properties that are not accessible to homogeneous systems at the same overall monomer concentration. In Model~II, instead, the gradient reorganizes monomers, fuel, and ions simultaneously, generating local combinations of concentrations associated with strongly enhanced supramolecular activity. 

Despite these differences, a common principle emerges from the two models: the local supramolecular dynamics is largely determined by the concentrations prevailing at each position along the gradient. The gradient therefore acts primarily by creating and spatially organizing local chemical environments that are not available to a homogeneous system with the same overall content of
self-assembling species, rather than by introducing a qualitatively different kinetic mechanism. These results identify chemical concentration gradients as a route to spatially control and locally activate supramolecular materials. More broadly, by applying the same approach to two mechanistically distinct models, we demonstrate that molecular simulations combined with controlled concentration gradients offer a general strategy to map how local chemical environments govern the structure and dynamics of spatially heterogeneous supramolecular systems, with potential relevance to both gradient-controlled experiments (such as micro- and nanofluidic control of self-assembly~\cite{sorrenti2016supramolecularchiral, rubiomartinez2016gradients, sevim2018microfluidic}) and the design of responsive materials.

\section*{Results}

We begin by comparing the two models under homogeneous ionic conditions to establish how their structure and dynamics respond to changes in ionic concentration. We then use $\cumd$ to impose stationary ionic gradients and examine how this affects the structure and dynamics of the assemblies at both global and local levels.

\begin{figure}[H]
  \centering
  \includegraphics[width=0.9\linewidth]{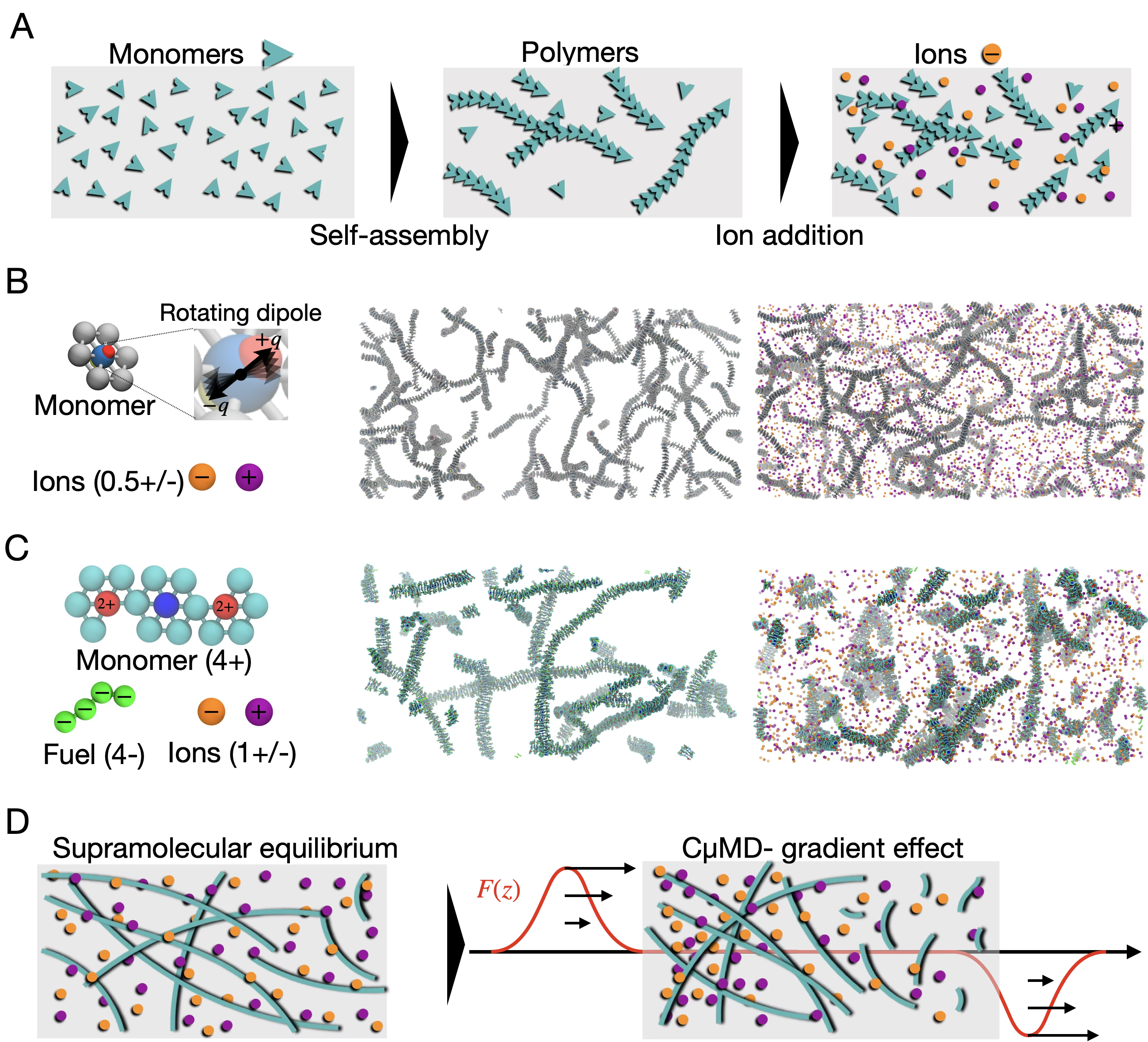}
  \caption{\textbf{Supramolecular polymerization models and simulation setup.}
    (A)~General concept: monomers self-assemble into supramolecular polymers, whose assembly state can be modulated by ionic species. (B)~Model~I, describing cooperative polymerization: monomers $\mi$ interact through Lennard--Jones (LJ) and rotating dipole--dipole interactions, while ionic species carry effective charge~$\pm0.5e$ and only weakly perturb the dipolar interactions. Snapshots show the assembled system without (center) and with (right) ions. (C)~Model~II, describing fuel-mediated polymerization: positively charged monomers $\mii$ (charge~$+4e$) are bridged by multivalent fuel molecules (charge~$-4e$), while ionic species (charge~$\pm1e$) compete with the fuel for monomer binding sites. Snapshots show the monomer--fuel system in the absence (center) and presence of ions (right). (D)~Left: homogeneous supramolecular equilibrium at uniform ionic concentration. Right: $\cumd$ setup, in which a position-dependent biasing force $F(z)$ maintains a steady ionic concentration gradient along $z$, sustaining the system in a spatially structured non-equilibrium steady state.}
  \label{fig:setup}
\end{figure}

We study two coarse-grained models that embody different self-assembly paradigms (Fig.~\ref{fig:setup}) and distinct modes of coupling to ionic species. Model~I describes cooperative polymerization of monomers, $\mi$, which self-assemble through Lennard--Jones (LJ) attraction and rotating dipole--dipole interactions~\cite{crippa2025nontrivial}; the simulated mobile ions carry effective charges $\pm0.5e$ and interact only weakly with the monomers, perturbing the dipolar interactions that drive cooperative self-assembly. Model~II instead mimics a fuel-mediated system in which positively charged monomers, $\mii$ (charge $+4e$), are bridged into assemblies by multivalent fuel molecules (each of total charge $-4e$)~\cite{Perego2021}; here, ionic species with charges $\pm1e$ compete with the fuel for monomer binding sites, resulting in a direct coupling of the ionic concentration to the self-assembly mechanism. 

To generate and sustain stationary ionic gradients, we employ the $\cumd$ technique~\cite{Karmakar2023}, which enforces prescribed ionic concentrations in two control regions, yielding an approximately linear steady-state profile along $z$, and maintains the system in a non-equilibrium steady state (Fig.~\ref{fig:setup}D). Full details of the molecular models and simulation protocols are provided in the Materials and Methods section.

\subsection*{Supramolecular response under homogeneous ionic conditions}
\begin{figure}[H]
  \centering
  \includegraphics[width=\linewidth]{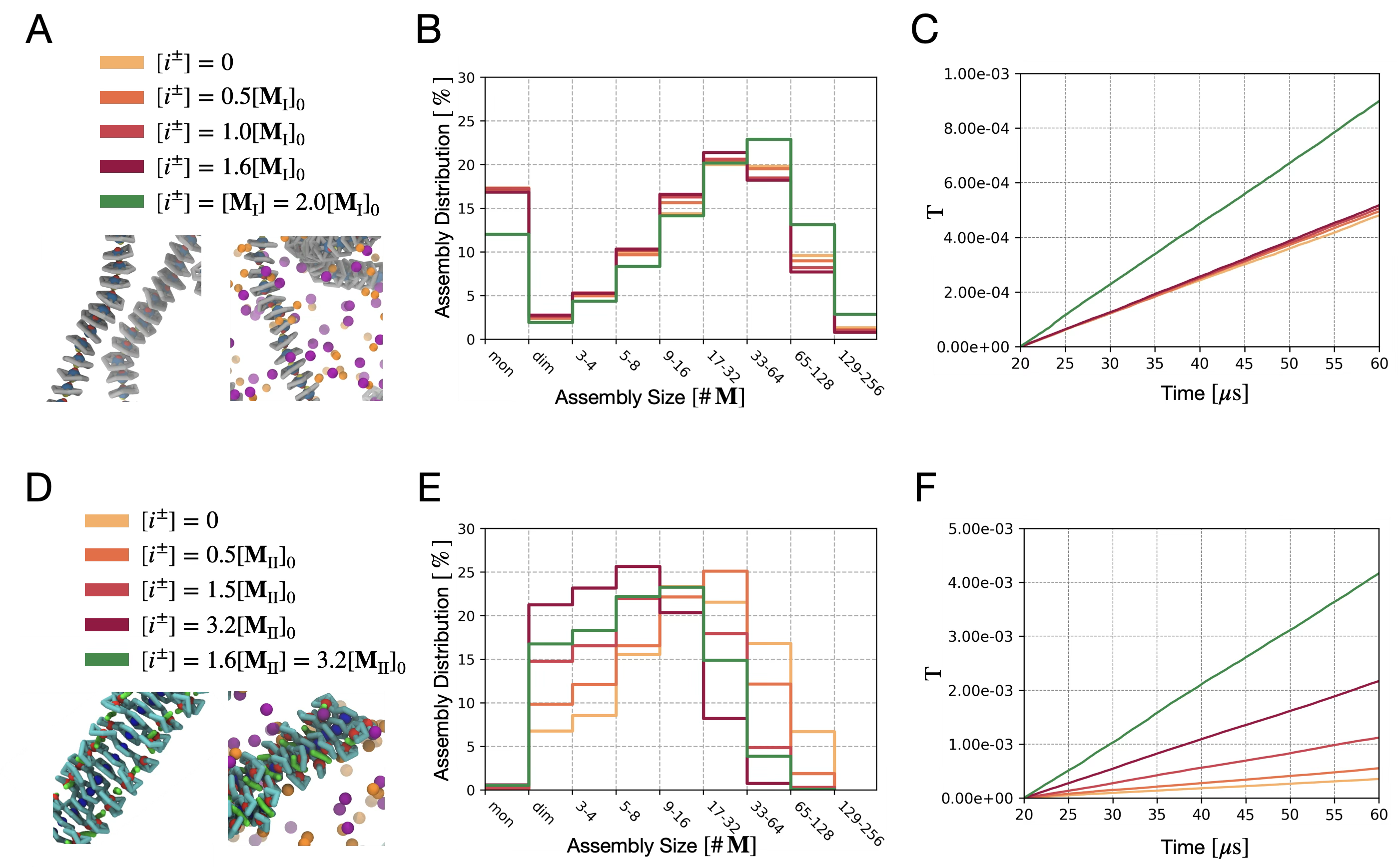}
    \caption{\textbf{Self-assembly under homogeneous ionic conditions.}
    (A,~D)~Color legends for Models~I and II, respectively, together with
    snapshots of aggregates in the absence and presence of ions.
    (B,~E)~Equilibrium assembly size distributions at different ionic
    concentrations $[i^\pm]$, and
    (C,~F)~molecular traffic $T$ as a function of simulation time, for Models I and II, respectively.}
  \label{fig:homogeneous}
\end{figure}

First, we establish a homogeneous reference by characterizing the self-assembly behavior of each model across a series of uniform ionic concentrations, $[i^\pm]$. At equilibrium, we quantify both structural and kinetic features of the supramolecular state via the distribution of assembly sizes and the molecular traffic~\cite{grant2011}, $T$, respectively; the latter is defined as the cumulative number of binding and unbinding events per monomer (Fig.~\ref{fig:homogeneous}). 

Model~I is almost insensitive to ionic concentration: across the full range of $[i^\pm]$ explored at fixed monomer concentration, $[\mi]_0$, both the equilibrium size distributions and the molecular traffic remain essentially unchanged, consistent with the weak coupling between ions and the dipolar interactions. By contrast, the half-volume run (see section Materials and Methods), in which both monomer and ion concentrations are doubled ($[i^\pm] = [\mi] = 2.0\,[\mi]_0$; green), displays a marked shift toward larger assemblies in the assembly size distribution and a clear increase in the molecular traffic. Since increasing $[i^\pm]$ alone up to $1.6\,[\mi]_0$ produces only minor changes, the effect on size distribution and traffic is likely due to the higher monomer concentration of the half-volume run.

Model~II, on the contrary, shows a two-fold response, with ions acting on both structure and dynamics. Increasing $[i^\pm]$ shifts the size distribution toward smaller assemblies, consistent with ions competing with the fuel for the monomer binding sites and perturbing the fuel-bridging mechanism. At the same time, the molecular traffic increases with $[i^\pm]$, as the smaller assemblies exchange monomers more readily. Thus, under homogeneous conditions, ionic concentration affects the structural and kinetic observables in opposite directions: increasing $[i^\pm]$ inhibits the formation of large assemblies while increasing the frequency of binding and unbinding events. 

Fig.~S1 provides a complementary characterization of the supramolecular equilibrium of the systems in homogeneous ionic conditions, through the average assembly size, the radial distribution functions of the monomer centers, and the size-dependent polymerization and depolymerization probabilities. These observables confirm the divergent behavior of the two models: Model~I is only weakly affected by variations in ionic concentration, whereas Model~II displays pronounced structural and kinetic responses, with shorter assemblies and higher molecular traffic as $[i^\pm]$ increases. This contrast provides the baseline against which the gradient-driven response of each system is assessed.

\subsection*{Supramolecular structure response upon ionic gradient}

\begin{figure}[H]
  \centering
  \includegraphics[width=0.9\linewidth]{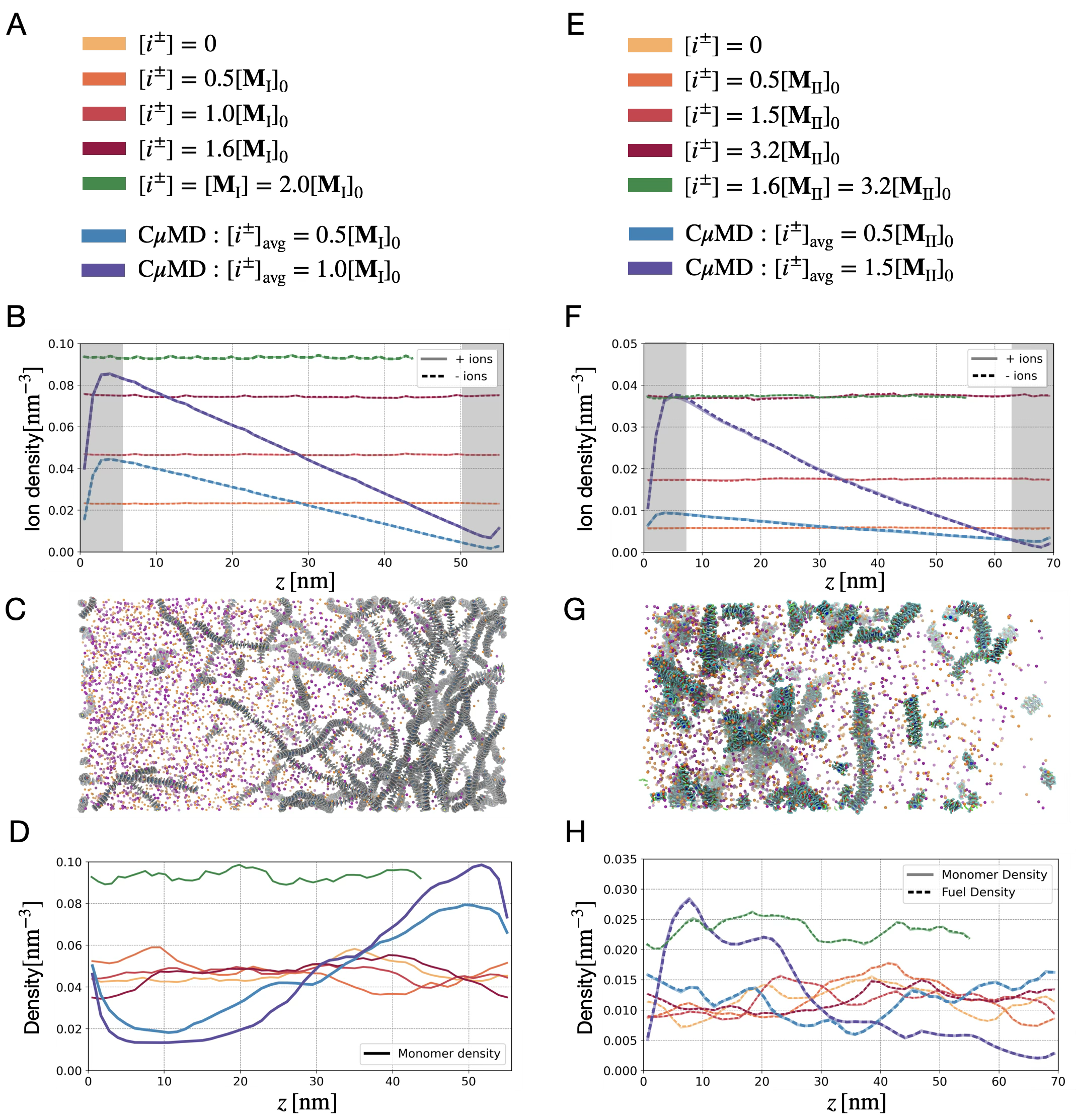}
  \caption{\textbf{Self-assembly coupled to ionic concentration gradients.}
    Left column: Model~I; right column: Model~II.
    (A,~E)~Color legend for the homogeneous (orange and green) and $\cumd$ gradient simulations
    (blue: weaker and purple: stronger gradient).
    (B,~F)~Ionic density profiles along the gradient direction $z$; solid and dashed
    lines denote positive and negative ionic species, respectively. $\cumd$ simulations display an approximately linear concentration profile between the control regions, whose locations are indicated by the gray shaded areas, whereas the homogeneous simulations are spatially uniform.
    (C,~G)~Representative snapshots of the supramolecular polymer configurations under the stronger ionic gradient (left: Model~I $[i^\pm]_{\mathrm{avg}}=1.0\,[\mi]_0$; right: Model~II $[i^\pm]_{\mathrm{avg}}=1.5\,[\mii]_0 $).
    (D,~H)~Density profiles of self-assembling species along~$z$; solid lines denote the monomer density, while dashed lines in~(H) indicate the fuel density.}
  \label{fig:spatial}
\end{figure}

We now impose an ionic concentration gradient and examine how it spatially reshapes the assemblies. The $\cumd$ protocol establishes a steady, linear ionic concentration profile along $z$, with the exception of the $z$-boundary, where the biasing force acts on the ions
(Fig.~\ref{fig:spatial}B,F). The concentration profiles remain stationary throughout the production runs. For each model, we explore a weaker and a stronger gradient condition (in blue and purple, respectively), generated by varying the total amount of ions and the $\cumd$ parameters (see Materials and Methods).

In Model~I, the gradient induces a response that was not anticipated by the homogeneous reference runs. Although the system showed only weak sensitivity to a uniform ionic perturbation, the gradient drives a marked redistribution of the monomers along $z$, with supramolecular assemblies becoming enriched in the low-ion region (Fig.~\ref{fig:spatial}D), phenomenologically reminiscent of gradient-driven transport phenomena previously reported~\cite{Shim2022_diffusiophoresis,Velegol2016,Illien2017_fuelled,Yadav2024_saltgradientinduced}. However, the resulting spatial pattern of assembly is a zero-current, thermodynamic redistribution rather than a kinetically driven transport: at steady state the monomer flux is null, and the distribution reflects the locally varying assembly equilibrium along the gradient. This behavior is consistent with a preferential-interaction mechanism~\cite{Knowles2011}, in which the ions would interact slightly less favorably with the assembled polymer than with free monomers. Within this interpretation, the local assembly would shift toward longer chains where the ionic activity is lowest. Even a weak preferential interaction, consistent with the limited ion sensitivity of Model~I, could produce substantial redistribution when integrated across the full gradient. A system that is only weakly affected by uniformly distributed ions can therefore develop a pronounced spatial response once a chemical gradient is established, with the stronger gradient producing a stronger redistribution.

As anticipated from the coupling observed under homogeneous conditions, in Model~II both the monomer and fuel density profiles develop a clear spatial structure along $z$ (Fig.~\ref{fig:spatial}H). The magnitude of the response again depends on the gradient strength: the stronger gradient produces clear spatial organization, whereas the weaker one barely perturbs the density profiles, suggesting that a minimum gradient strength may be required to activate a substantial response. Because the fuel is almost entirely incorporated into the assemblies, its density profile provides a proxy for the spatial distribution of the supramolecular polymers. The co-localization of monomer and fuel in the high ion concentration region therefore reflects the preferential accumulation of the assembled $\mii$--fuel polymers there. The gradient generates local combinations of $[\mii]$, $[\mathrm{fuel}]$, and $[i^\pm]$  in which all three species are simultaneously present at high concentrations. Under homogeneous conditions, such combinations could be reached only by increasing the monomer, fuel, and ion concentrations, whereas here they emerge through spatial redistribution driven solely by the modulation of the ionic environment.

\subsection*{Assembly distributions and molecular traffic under the gradient}

\begin{figure}[H]
  \centering
  \includegraphics[width=\linewidth]{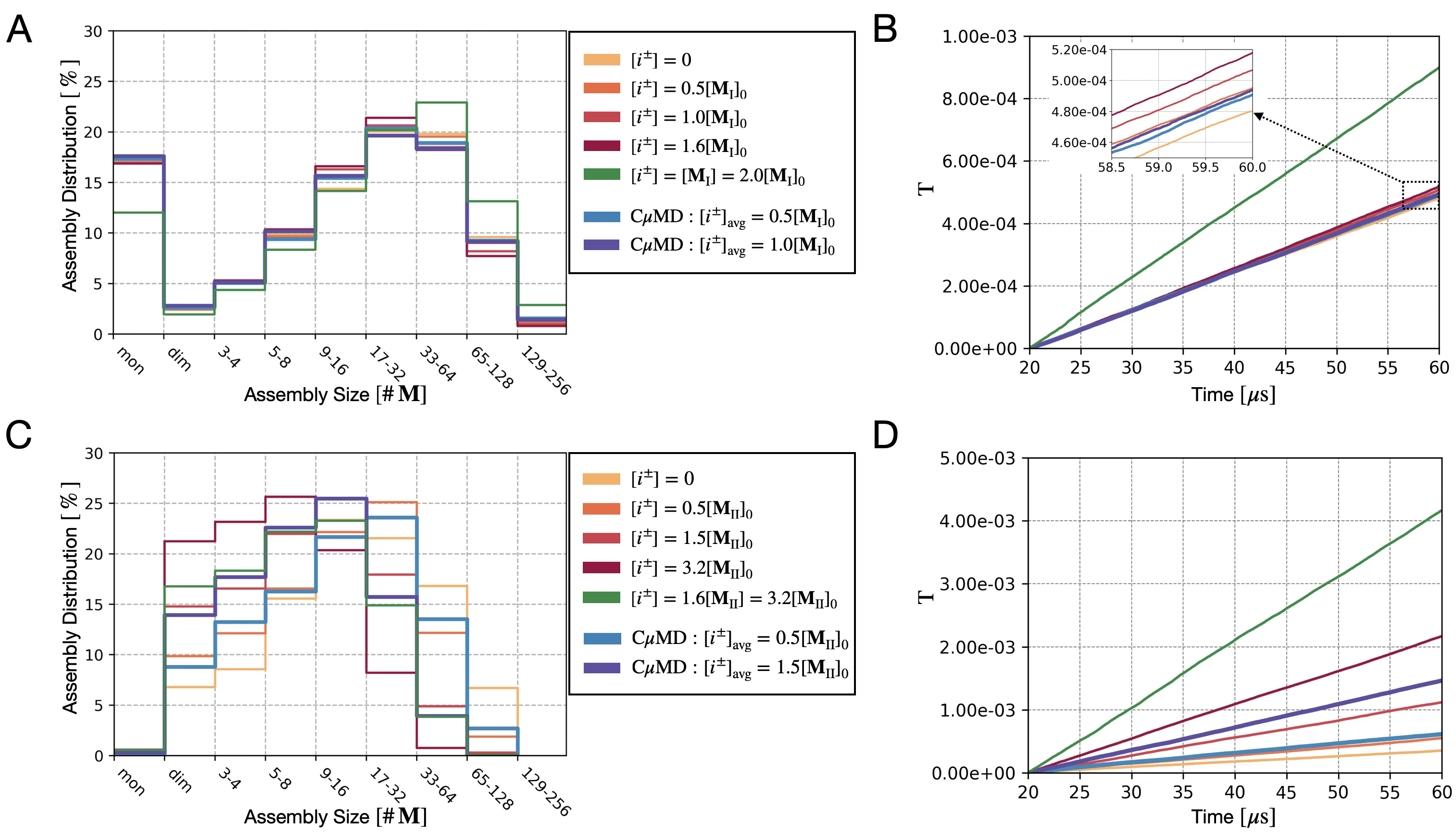}
  \caption{\textbf{Assembly size distributions and molecular traffic under
    ionic gradient conditions compared to the homogeneous references.}
    (A,~C)~Steady-state assembly size distributions for Model~I~and Model~II, respectively.
    (B,~D)~Molecular traffic $T$ as a function of the simulation time; the inset in (B) magnifies the late-time region.}
  \label{fig:gradient_assembly}
\end{figure}

We next examine how this spatial structuring is reflected in global observables of supramolecular structure and dynamics. For both models, the global assembly size distributions obtained in the $\cumd$ simulations are consistent with the homogeneous reference distributions at comparable
average ionic concentrations (Fig.~\ref{fig:gradient_assembly}A,C). 

As with the assembly size distribution, the molecular traffic of Model~I under gradient conditions remains consistent with that of the homogeneous reference at the same average ionic concentration
(Fig.~\ref{fig:gradient_assembly}B, inset). As expected from the weak ionic coupling, the gradient neither suppresses nor enhances the overall exchange dynamics relative to an equivalent well-mixed system with the same average composition. This consistency is also reflected in the average assembly size, the radial distribution functions, and the size-dependent probabilities of polymerization and depolymerization reported in Fig.~S2.

By contrast, Model~II exhibits a clear dynamical response to the ionic gradient. Its molecular traffic is enhanced relative to the homogeneous reference with the same global composition: the simulation with the stronger gradient exceeds the homogeneous reference at $[i^\pm]=1.5\,[\mii]_0$, while under the weaker gradient it exceeds the corresponding homogeneous reference at $0.5\,[\mii]_0$ (Fig.~\ref{fig:gradient_assembly}D). This enhancement is consistent with the co-localization described above. In regions where the ionic concentration is highest, supramolecular polymers are also locally enriched and experience local compositions that, in homogeneous systems, are associated with higher molecular traffic. The co-localization of polymers and ions therefore provides a plausible explanation for the enhanced global traffic.

These global quantities provide only a partial (averaged) description, as system-wide averaging conceals the spatial heterogeneity evident in Fig.~\ref{fig:spatial}. A single $\cumd$ system, once the steady state is reached, hosts regions of very different local character---from monomer- and polymer-depleted zones to regions with high density of self-aggregated matter---a heterogeneity that the global distribution collapses into a single averaged profile. To identify the mechanistic origin of these behaviors, we move our analysis to local (spatially resolved) features.

\subsection*{Local supramolecular activity}

\begin{figure}[H]
  \centering
  \includegraphics[width=\linewidth]{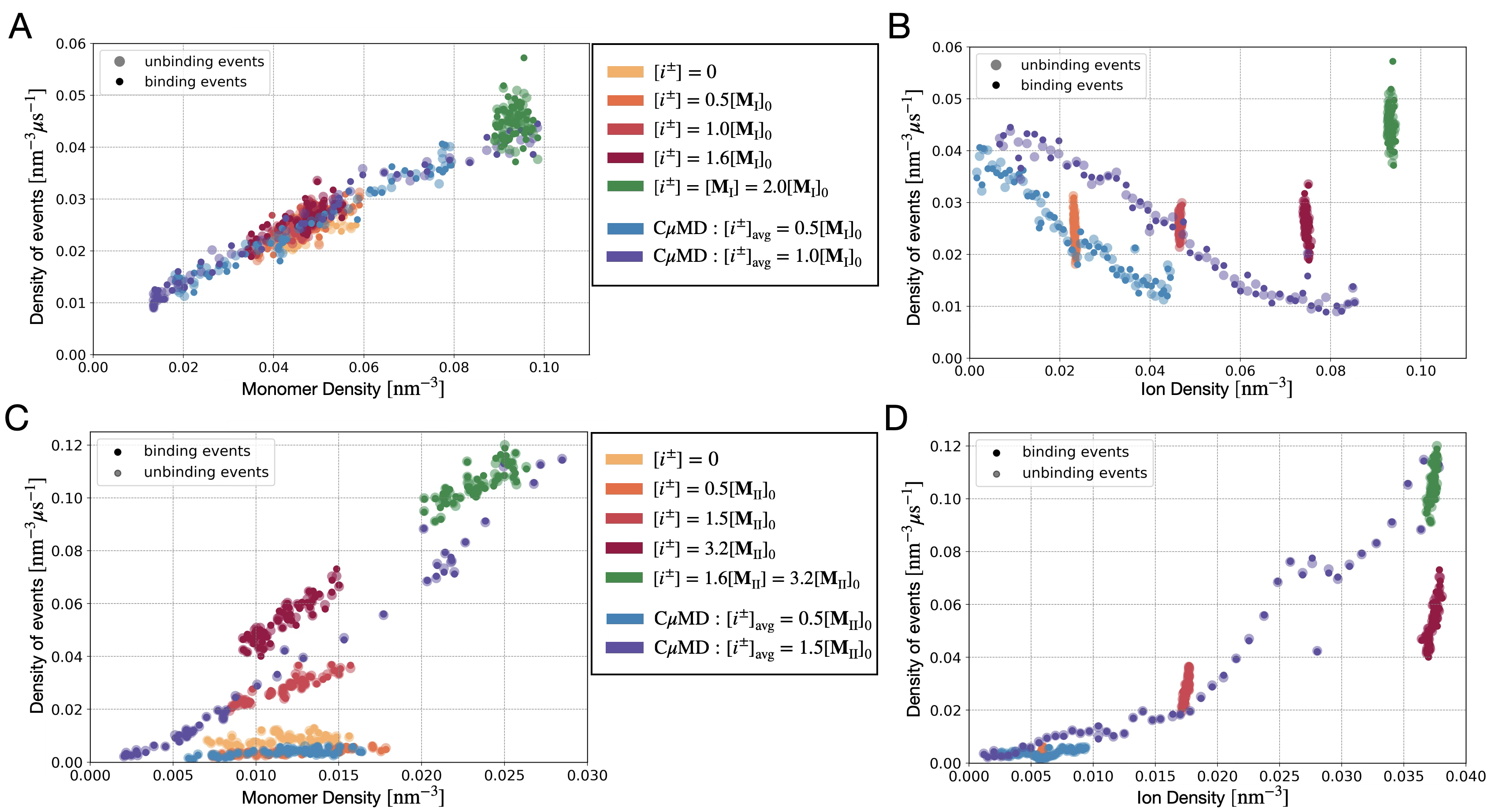}
  \caption{\textbf{Supramolecular activity in homogeneous and ionic gradient simulations.}
   Density of binding (small, opaque circles) and unbinding (larger, transparent circles, same color as binding) events as a function of local
   monomer density (A, C) and local ion density (B, D) for Model I (A, B) and Model II (C, D). Color coding as in Figs.~\ref{fig:spatial}--\ref{fig:gradient_assembly}.}
  \label{fig:activity}
\end{figure}

To rationalize the gradient-induced responses, we now resolve the assembly dynamics locally. We define the local supramolecular activity as the spatially resolved density of binding and unbinding events between monomers along $z$ (Fig.~S3), and examine its dependence on the local monomer and ion density (Fig.~\ref{fig:activity}). This analysis tests whether the dynamical response at each position along the gradient is governed by the locally prevailing concentrations.

For Model~I, the local analysis shows that the data from both homogeneous and gradient simulations collapse onto a single master curve when the local activity is plotted against the local monomer density (Fig.~\ref{fig:activity}A), while no equivalent collapse is observed against the local ion density (Fig.~\ref{fig:activity}B). Ions therefore modulate assembly primarily through their indirect effect on local monomer availability, rather than through a direct influence on binding. By redistributing monomers, the gradient creates regions of high local monomer density and enhanced activity that are not accessible to a homogeneous system without changing the global monomer content.

Model~II again exhibits a different local response. Because ions are directly coupled to the assembly mechanism, homogeneous simulations do not collapse onto a single master curve; instead, each ionic concentration defines a distinct activity curve (Fig.~\ref{fig:activity}C,D). The condition with the stronger gradient does not follow any single homogeneous curve but spans the family of homogeneous responses as both $[\mii]$ and $[i^\pm]$ vary together along $z$. As supramolecular polymers accumulate where the ionic concentration is higher, the most active regions combine high local polymer density with high local ion density. When plotted against the local ion density alone (Fig.~\ref{fig:activity}D), the gradient points approach the high-monomer-density branch of the homogeneous simulations, indicating that activity depends on the joint local composition $([\mii](z),[\mathrm{fuel}](z),[i^\pm](z))$, rather than on ionic concentration alone. This behavior becomes evident under the stronger gradient, whereas the weaker gradient produces only minor deviations from the homogeneous response, again revealing a pronounced sensitivity to gradient intensity. 

Consistently, Fig.~S3 shows that the spatial distribution of binding and unbinding events mainly follows the local monomer density in Model~I, whereas in Model~II it follows the product of the local monomer and ion densities.
The different behavior of the two models can therefore be traced to the way ionic species couple to the assembly mechanism. 
In Model~I, ions only weakly perturb the interactions that stabilize supramolecular polymers, so that local activity is  controlled mainly by the local monomer density: monomer enrichment increases the frequency of intermolecular encounters, and thus the rate of binding events, while the stability of existing monomer--monomer contacts is only weakly affected by the local ionic concentration. In Model~II, by contrast, ions compete directly with the multivalent fuel for monomer binding sites, so changes in local ionic concentration affect the fuel-mediated association mechanism itself. Local activity in Model II is thus jointly controlled by molecular encounter rates and the local balance between monomers, fuel, and ions. 

Despite these mechanistic differences, both models point to a common principle: the local supramolecular activity is largely determined by the concentrations prevailing at each position along the gradient, with no detectable residual dependence on the gradient itself. Each region therefore behaves kinetically like a homogeneous subsystem at constant local composition, and the local densities of binding and unbinding events remain closely balanced. These observations are consistent with a local quasi-equilibrium description of the simulated non-equilibrium steady state.

\section*{Discussion and Conclusions}

Using coarse-grained molecular dynamics combined with $\cumd$, we have characterized how ionic concentration gradients can shape the structure and dynamics of two models representing fundamentally different classes of supramolecular polymer systems. Model~I mimics the self-assembly behavior of cooperative supramolecular polymers, whereas Model~II reproduces the dynamics of fuel-mediated polymerization. Moreover, these two models couple differently to the ionic species. Under homogeneous conditions, Model~I is only weakly sensitive to ionic concentration, with assembly governed primarily by monomer density. By contrast, in Model~II, increasing the ionic concentration simultaneously reduces the average assembly size and increases molecular traffic. 

When an ionic concentration gradient is imposed, Model~I undergoes a marked spatial redistribution, with polymeric chains accumulating where the ionic concentration is lowest, revealing that a weak ionic sensitivity in a homogeneous environment does not preclude a strong gradient-driven response. Although this redistribution leaves the global structural and dynamical observables essentially unchanged, it creates local regions of high monomer concentration and, consequently, enhanced supramolecular activity that cannot be accessed by a homogeneous system with the same global monomer content. Model~II responds differently: its monomer and fuel distributions co-localize where the ionic concentration is highest, consistent with the direct coupling between ionic concentration and assembly. At the global level, this co-localization enhances molecular traffic relative to the homogeneous reference at the same mean composition. Locally, the simultaneous enrichment of monomers, fuel, and ions generates levels of high supramolecular activity that, again, cannot be attained in homogeneous conditions without increasing the total content of self-assembling species.

Although ionic gradients trigger an opposite spatial redistribution in the two models, a common governing principle emerges. In both systems, the supramolecular activity at a given position is largely determined by the concentrations prevailing locally, consistent with a state of \textit{local quasi-equilibrium}. Under this interpretation, each region along the gradient locally resembles a homogeneous reference system at the same local concentrations of the relevant species. A possible physical basis for this behavior is that the $\cumd$ biasing force acts directly on the ions, while the monomers respond indirectly and relax to the zero-current distribution set by the local free energy landscape. This locally balanced character distinguishes our systems from dissipative assemblies driven by synthetic reaction cycles, in which fuel consumption sustains reaction fluxes and  breaks the detailed balance globally~\cite{Ragazzon2018,Riess2020}. A practical consequence is that a single gradient simulation simultaneously samples a continuous range of local compositions, thereby providing a map between local conditions and supramolecular activity that would otherwise require a full series of homogeneous simulations.

It should be noted that the gradients considered in this study extend over a nanometer length scale, whereas experimentally investigated ionic gradients, and gradient-directed self-organization processes,  typically extend over micrometer-to-millimeter distances~\cite{Nguindjel2022,Lovrak2017}. Nevertheless, nanoscale ionic heterogeneity is also physically relevant. For instance, in biological systems, Ca$^{2+}$ nanodomains develop around individual ion channels and control local signaling processes~\cite{Ramachandran2022_nanodomains}, while transmembrane proton gradients can regulate the spatial organization of bacterial membrane domains~\cite{prithviraj2025_protongradient}. Strong ionic heterogeneity also arises in engineered confined systems, including through ion-concentration polarization at micro- and nano-fluidic interfaces~\cite{Ouyang2018_deciphering} and transient salt gradients driving diffusiophoretic transport in dead-end pores~\cite{migacz2024_enhanceddiffusiophoresis}.               

Furthermore, the local quasi-equilibrium description observed in our simulations reduces the importance of the overall gradient length scale. Since the local supramolecular activity is largely determined by the concentrations prevailing at each position, the same local response could, in principle, occur within a larger system in which comparable concentration variations are distributed over a greater distance. Gentler micrometer-to-millimeter gradients may also be more compatible with a local-equilibrium description, because the concentrations vary more slowly in space. Our nanoscale simulations therefore provide molecular-level insights into local responses that may also be relevant to gradients extending  well beyond the simulation cell.

Taken together, these results identify chemical gradients as a powerful handle for controlling supramolecular materials. Systems that are only weakly responsive to homogeneous ionic perturbation, such as Model~I, may nevertheless undergo pronounced spatial reorganization when exposed to a sufficiently strong ionic gradient. On the other hand, systems with a stronger homogeneous response, such as Model~II, offer a complementary opportunity: the gradient generates local combinations of monomer, fuel, and ion concentrations that cannot be reproduced homogeneously without changing the global composition, thereby promoting the gradient-controlled co-localization of supramolecular species explored by recent experiments~\cite{Pryanka2024_ATPgradient}. These observations suggest a possible design principle: the direction of the homogeneous ionic response may help predict how a gradient will spatially organize a given system, even when the magnitude of that response does not anticipate the strength of the gradient-driven redistribution, and could therefore guide the design of specific supramolecular patterns and activated regions. 

Several extensions follow naturally: multicomponent or temporally modulated gradients could enable more elaborate spatial and dynamical control of supramolecular assemblies; coupling gradient control to active matter or non-equilibrium fuel cycles could generate more complex driven behaviors; and studies at larger, experimentally accessible length scales could connect the mechanism identified here to functional materials. Overall, chemical gradients can provide a versatile route toward the rational, out-of-equilibrium design of supramolecular matter capable of reproducing key features of living assemblies that form and function in spatially heterogeneous chemical environments.

\section*{Materials and Methods}

\subsection*{Coarse-grained models}

\subsubsection*{Model~I: cooperative supramolecular polymerization}

Model~I is the cooperative supramolecular polymer model introduced in Ref.~\cite{crippa2025nontrivial}, adopted here without modification. It is formed by the self-assembly of hexagonal monomers, $\mi$. Each monomer consists of nine interacting beads (see Fig.~\ref{fig:setup}B): six peripheral shielding beads at the vertices of a hexagon and a central core bead (all with $\sigma = 0.47$\,\si{\nano\meter} and mass $72$\,\si{\atomicmassunit}) are held together by harmonic bonds connecting both nearest neighbors (force constant $k = 20\,000$\,\si{\kilo\joule\per\mole\per\nano\meter\squared}, equilibrium length $l_0 = 0.47$\,\si{\nano\meter}) and cross-diagonal vertex beads ($k = 15\,000$\,\si{\kilo\joule\per\mole\per\nano\meter\squared}, $l_0 = 0.94$\,\si{\nano\meter}). 

Nonbonded interactions between core beads are described by a Lennard--Jones (LJ) potential ($\sigma = 0.47$\,\si{\nano\meter},
$\varepsilon = 20$\,\si{\kilo\joule\per\mole}), while shielding beads interact weakly with all other beads ($\varepsilon = 0.2$\,\si{\kilo\joule\per\mole}) and act primarily as excluded volume, guiding monomer stacking along the symmetry axis. The two remaining beads form a freely rotating dipole, attached to the core bead, that introduces the cooperative character of the model. The dipole beads bear effective charges $q = \pm 1.4$\,\si{\elementarycharge} and are both constrained at a fixed distance of $0.14$\,\si{\nano\meter} from the core-bead center, forming two vectors. These two vectors are maintained antiparallel by a harmonic angular potential (equilibrium angle $180^\circ$, force constant $1500$\,\si{\kilo\joule\per\mole\per\radian\squared}), resulting in a dipole vector $0.28$\,\si{\nano\meter} long and an effective dipole moment of $0.392$\,\si{\elementarycharge\nano\meter} ($\approx 18.8$~D). The dipole beads carry one third of the mass of a regular bead ($24$\,\si{\atomicmassunit}) and interact exclusively through Coulomb interactions, with no excluded-volume contribution. As shown in Ref.~\cite{crippa2025nontrivial}, dimerization in this model is entropically penalized because two freely diffusing monomers must meet with suitably aligned dipoles to form a stable interaction. In longer oligomers, by contrast, the dipoles at the ends are already preferentially aligned along the direction orthogonal to the monomer plane, facilitating the binding of further monomers or polymer ends. The resulting difference between initial association and subsequent fiber elongation gives rise to a nucleation-like mechanism underlying the cooperative polymerization.

In this model, mobile ions in solution carry effective charges $q = \pm 0.5$\,\si{\elementarycharge} and have a mass of $47$\,\si{\atomicmassunit}. They interact with all beads through LJ potentials ($\sigma = 0.47$\,\si{\nano\meter}, $\varepsilon = 0.2$\,\si{\kilo\joule\per\mole}).%

All simulations of Model~I are performed in a rectangular cell of dimensions $27.8 \times 27.8 \times 55.6$\,\si{\nano\meter\cubed} ($N = 2000$ monomers, concentration $\approx\!77.3$\,mM) with periodic boundary conditions in all directions. The simulation $[i^\pm] = [\mi] = 2.0\,[\mi]_0$
 is the only exception, as it was performed in a rectangular cell of $22.1\times22.1\times44.1$\,\si{\nano\meter\cubed}, having half the volume of all other cases.

\subsubsection*{Model~II: fuel-mediated supramolecular polymerization}

Model~II is based on the CG model introduced in Ref.~\cite{Perego2021}, which was developed as a minimal representation of the ATP-responsive supramolecular polymer reported experimentally in Ref.~\cite{Mishra2018}. In the experimental system, monomers containing an oligo($p$-phenylenevinylene) core functionalized at both ends with Zn(II)-dipicolylamine receptors undergo cooperative polymerization upon ATP binding, the latter acting as a multivalent bridge between monomers. The CG model retains this electrostatic bridging mechanism by representing the monomers, $\mii$, as positively charged species with total charge $+4\,$\si{\elementarycharge}, and the fuel molecules as multivalent linkers with total charge $-4\,$\si{\elementarycharge}. In detail, each $\mii$ monomer consists of 17 beads interacting via Coulomb and LJ potentials (see Fig.~\ref{fig:setup}C): one central core bead and 14 screening beads (mass $72$\,\si{\atomicmassunit} each), and two charged peripheral beads (charge $+2\,\si{\elementarycharge}$ and mass $47$\,\si{\atomicmassunit} each). Nearest-neighbor harmonic bonds have force constant $k = 15\,000$\,\si{\kilo\joule\per\mole\per\nano\meter\squared} and equilibrium length $l_0 = 0.47$\,\si{\nano\meter}; second-neighbor bonds ($k = 10\,000$\,\si{\kilo\joule\per\mole\per\nano\meter\squared}) maintain monomer planarity. Each fuel molecule consists of four linearly connected beads of mass $47$\,\si{\atomicmassunit} each and charge $-1\,\si{\elementarycharge}$ (total $-4\,\si{\elementarycharge}$), held together by harmonic bonds ($k = 15\,000$\,\si{\kilo\joule\per\mole\per\nano\meter\squared}, $l_0 = 0.30$\,\si{\nano\meter}) and a cosine-harmonic angular potential (equilibrium angle $180^\circ$, force constant $50$\,\si{\kilo\joule\per\mole}). Fuel molecules bind to $\mii$ monomers and bridge them into supramolecular polymers via electrostatic attraction; monovalent salt ions (charge $\pm 1\,\si{\elementarycharge}$, mass $47$\,\si{\atomicmassunit}) compete with fuel binding.

Nonbonded interactions are described by LJ potentials with $\sigma = 0.47$\,\si{\nano\meter} for all bead pairs; the core bead self-interaction has $\varepsilon = 30$\,\si{\kilo\joule\per\mole}, while all other pairs---including fuel and salt ion beads---use a weak $\varepsilon = 0.2$\,\si{\kilo\joule\per\mole}. %

Simulations of Model~II are performed in a box of $35 \times 35 \times 70$\,\si{\nano\meter\cubed} ($N = 1000$ monomers and $N_\mathrm{fuel}=1000$ fuel molecules, monomer concentration $\approx\!19.4$\,mM) with periodic boundary conditions in all directions. The $[\mii] = 2.0\,[\mii]_0$ simulation is the only exception, as it was performed in a rectangular cell of $27.8\times27.8\times55.6$\,\si{\nano\meter\cubed}, having half the volume of all other cases.  

\subsection*{CG-MD simulation protocol}\label{sub:CGMD}

All simulations are performed with the GROMACS software~\cite{Hess2008} in NVT conditions using implicit solvent Langevin dynamics. The stochastic dynamics (\texttt{sd}) integrator is used with a time step $\Delta t = 20$\,\si{\femto\second}, inverse friction constant $\tau_c = 0.1$\,\si{\pico\second}, and temperature $T = 300$\,\si{\kelvin}. This choice places the dynamics in the strongly overdamped regime, suppressing inertial effects. 

Coulomb interactions are included between all charged species, using a relative permittivity $\varepsilon_\mathrm{r} = 15$ for dielectric screening, consistent with the implicit solvent,  dry Martini style,~\cite{Arnarez2015} representation. All nonbonded interactions (Coulomb included) are truncated and shifted to zero at a cutoff $r_c = 1.1$\,\si{\nano\meter}.

Homogeneous reference simulations are performed at several ionic concentrations $[i^\pm]$ (Tables~\ref{tab:simI}--\ref{tab:simII}). A half-volume run is included for each model, in which all box dimensions are scaled by $2^{-1/3}$ to halve the simulation volume, doubling the monomer concentration (and the fuel concentration for Model~II) with respect to the other runs. The ionic concentration is chosen independently: $[i^\pm]=[\mi]$ for Model~I and $[i^\pm]=1.6\,[\mii]$ for Model~II. local-Each system is initialized from randomly distributed configurations and run for a total of $60$\,\si{\micro\second}; equilibrium is confirmed by monitoring the average assembly size as a function of time (Fig.~S1). The last 40\,\si{\micro\second} are used for the analysis, except for the spatial profiles (Fig.~\ref{fig:spatial}), radial distribution functions (Fig.~S1), and local activity analyses (Figs.~\ref{fig:activity} and S3), which are all computed over the final 20\,\si{\micro\second} of the trajectories. In the Model II simulation at $[\mii] = 2.0\,[\mii]_0$, the total run lasted for $100$\,\si{\micro\second}, as reaching supramolecular equilibrium required longer time, due to the higher overall density. For this specific simulation, all the analyses are performed between $60$ and $100$\,\si{\micro\second} of the trajectory.

\begin{table}[h]
\centering
\caption{\textbf{Model~I simulation runs.}
  $N_{i^\pm}$: number of positive/negative ions per species; $[\mi]_0 = 77.3$\,mM.
  All concentrations in mM.
  \textsuperscript{$a$}Half-volume run: all box dimensions scaled by $2^{-1/3}$.}
\label{tab:simI}
\begin{tabular}{clrrrr}
\toprule
 & & $N_{\mi}$ & $N_{i^\pm}$ & $[\mi]$ & $[i^\pm]$ \\
\midrule
{\large$\textcolor{colA}{\bullet}$} & $[i^\pm]=0$                                        & 2000 &    0 &  77.3 &   0.0 \\
{\large$\textcolor{colB}{\bullet}$} & $[i^\pm]=0.5\,[\mi]_0$                      & 2000 & 1000 &  77.3 &  38.6 \\
{\large$\textcolor{colC}{\bullet}$} & $[i^\pm]=1.0\,[\mi]_0$                      & 2000 & 2000 &  77.3 &  77.3 \\
{\large$\textcolor{colD}{\bullet}$} & $[i^\pm]=1.6\,[\mi]_0$                      & 2000 & 3200 &  77.3 & 123.7 \\
{\large$\textcolor{colE}{\bullet}$} & $[i^\pm]=[\mi]=2.0\,[\mi]_0$\textsuperscript{$a$} & 2000 & 2000 & 154.2 & 154.2 \\
{\large$\textcolor{colF}{\bullet}$} & C$\mu$MD,\ $[i^\pm]_{\mathrm{avg}}=0.5\,[\mi]_0$ & 2000 & 1000 &  77.3 & --- \\
{\large$\textcolor{colG}{\bullet}$} & C$\mu$MD,\ $[i^\pm]_{\mathrm{avg}}=1.0\,[\mi]_0$ & 2000 & 2000 &  77.3 & --- \\
\bottomrule
\end{tabular}
\end{table}

\begin{table}[h]
\centering
\caption{\textbf{Model~II simulation runs.}
  $N_{i^\pm}$: number of positive/negative ions per species; $[\mii]_0=[\mathrm{fuel}]_0=19.4$\,mM.
  All concentrations in mM.
  \textsuperscript{$a$}Half-volume run: all box dimensions scaled by $2^{-1/3}$.}
\label{tab:simII}
\begin{tabular}{clcrcr}
\toprule
 & & $N_{\mii}\!=\!N_\mathrm{fuel}$ & $N_{i^\pm}$ & $[\mii]\!=\![\mathrm{fuel}]$ & $[i^\pm]$ \\
\midrule
{\large$\textcolor{colA}{\bullet}$} & $[i^\pm]=0$                                                           & 1000 &    0 & 19.4 &  0.0 \\
{\large$\textcolor{colB}{\bullet}$} & $[i^\pm]=0.5\,[\mii]_0$                                        & 1000 &  500 & 19.4 &  9.7 \\
{\large$\textcolor{colC}{\bullet}$} & $[i^\pm]=1.5\,[\mii]_0$                                        & 1000 & 1500 & 19.4 & 29.0 \\
{\large$\textcolor{colD}{\bullet}$} & $[i^\pm]=3.2\,[\mii]_0$                                        & 1000 & 3200 & 19.4 & 62.0 \\
{\large$\textcolor{colE}{\bullet}$} & $[i^\pm]=1.6\,[\mii]=3.2\,[\mii]_0$\textsuperscript{$a$} & 1000 & 1600 & 38.8 & 61.8 \\
{\large$\textcolor{colF}{\bullet}$} & C$\mu$MD,\ $[i^\pm]_{\mathrm{avg}}=0.5\,[\mii]_0$               & 1000 &  500 & 19.4 & --- \\
{\large$\textcolor{colG}{\bullet}$} & C$\mu$MD,\ $[i^\pm]_{\mathrm{avg}}=1.5\,[\mii]_0$               & 1000 & 1500 & 19.4 & --- \\
\bottomrule
\end{tabular}
\end{table}

\subsection*{C$\mu$MD simulations}
To impose and maintain a steady ionic concentration gradient along $z$, we use the constant chemical potential molecular dynamics ($\cumd$) technique~\cite{Perego2015,Ozcan2017,Karmakar2023}. Two control regions, CR$^{\mathrm{high}}$ and CR$^{\mathrm{low}}$, each of thickness $0.1\,L_z$ (where $L_z$ is the box side length along $z$) are placed at opposite sides of the simulation box. The local ionic concentrations measured in these regions are compared with the prescribed target values, $[i^\pm]_\mathrm{high}$ in CR$^{\mathrm{high}}$ and $[i^\pm]_\mathrm{low}$ in CR$^{\mathrm{low}}$. A position-dependent biasing force,  $F(z)$, is adjusted on the fly to drive the concentrations toward their respective targets. Because ions continuously diffuse away from the control regions, the feedback force remains permanently active, sustaining the ionic gradient throughout the simulation. The target concentrations listed in Table~\ref{tab:gradients} are prescribed values; the concentrations actually attained in the control regions are lower, which is why the effective gradient intensity $g$ is used to characterize the gradient strength. $F(z)$ is localized near the periodic boundary along $z$ through a bell-shaped function with characteristic width $0.01\,L_{z}$. 

For each model, we simulate two gradient conditions, referred to as weak and strong (Table~\ref{tab:gradients}). For Model~I, the same target concentrations are used in both cases, $[i^\pm]_\mathrm{high}=0.125$\,\si{\nano\meter^{-3}} in CR$^{\mathrm{high}}$ and $[i^\pm]_\mathrm{low}=0$ in CR$^{\mathrm{low}}$. The two conditions differ only in the number of ions per species, with $N_{i^\pm}=1000$ for the weak gradient and $N_{i^\pm}=2000$ for the strong gradient. For Model~II, both the number of ions and the target concentration in CR$^{\mathrm{high}}$ are varied. The weak gradient is generated using $N_{i^\pm}=500$ and $[i^\pm]_\mathrm{high}=0.060$\,\si{\nano\meter^{-3}}, whereas the strong gradient uses $N_{i^\pm}=1500$ and
$[i^\pm]_\mathrm{high}=0.175$\,\si{\nano\meter^{-3}}; $[i^\pm]_\mathrm{low}=0$ in both cases. Because target concentrations are not necessarily reached in the stationary state, the effective gradient intensity is quantified directly from the simulated ionic concentration profiles as $g=\left|\frac{d[i^\pm]}{dz}\right|$. The slope is obtained by fitting the approximately linear portion of the steady-state profile in the unforced region, excluding the control regions and the nonlinear boundary layers adjacent to them. The resulting gradient intensities are reported in Table~\ref{tab:gradients}.

\begin{table}[h]
\centering
\caption{\textbf{Ionic gradient conditions.}
$N_{i^\pm}$ is the number of ions per species. The effective gradient intensity $g$ is estimated from a linear fit of the steady-state ionic concentration profile within the approximately linear, unforced region, excluding the control regions and the adjacent nonlinear boundary layers.}
\label{tab:gradients}
\begin{tabular}{llrrrr}
\toprule
Model &
Condition &
$N_{i^\pm}$ &
$[i^\pm]_\mathrm{high}$ &
$[i^\pm]_\mathrm{low}$ &
$g$ \\
&
&
&
(\si{\nano\meter^{-3}}) &
(\si{\nano\meter^{-3}}) &
(\si{\nano\meter^{-4}}) \\
\midrule
I  & weak   & 1000 & 0.125 & 0 & $8.9\times10^{-4}$ \\
I  & strong & 2000 & 0.125 & 0 & $1.6\times10^{-3}$ \\
II & weak   &  500 & 0.060 & 0 & $1.1\times10^{-4}$ \\
II & strong & 1500 & 0.175 & 0 & $6.1\times10^{-4}$ \\
\bottomrule
\end{tabular}
\end{table}

All $\cumd$ simulations are initialized from randomly distributed configurations and run for a total of $60$\,\si{\micro\second}. Steady-state conditions are assessed by monitoring the average assembly size as a function of time (Fig.~S2). The last 40\,\si{\micro\second} are used for analysis, except for the spatial profiles (Fig.~\ref{fig:spatial}), for the radial distribution functions (Fig.~S2), and for the local-activity analyses (Figs.~\ref{fig:activity} and S3), which are all computed over the final 20\,\si{\micro\second} of the trajectories. All reported observables, including the global assembly size distributions and molecular traffic, are computed over the full simulation box, including the control regions. The biasing force is confined to a narrow boundary slab of characteristic width $0.01\,L_z$ and acts directly on the ions. Moreover, the local binding and unbinding activity measured in the control region bins remains consistent with the corresponding homogeneous reference data (Figs.~\ref{fig:activity} and S3), suggesting that the biasing protocol does not produce a detectable alteration of the local supramolecular kinetics.

\subsection*{Analysis}

All analyses are performed using in-house Python scripts and the MDAnalysis package~\cite{Gowers2016,MichaudAgrawal2011}. All simulation snapshots reported are rendered using VMD~\cite{Humphrey1996}.

\subsubsection*{Assembly identification and structural characterization}

Supramolecular binding and unbinding events are identified using a two-cutoff criterion. A bond between two monomers is considered to form when their core-to-core distance falls below a binding cutoff $r_b = 0.6$\,\si{\nano\meter}, whereas an existing bond is considered broken when the same distance exceeds the unbinding cutoff $r_u = 0.9$\,\si{\nano\meter}. The resulting hysteresis suppresses spurious events caused by short-lived distance fluctuations. For Model~II, binding and unbinding events between fuel beads and the charged beads of $\mii$ are tracked separately using, respectively, the following cutoffs:  $r'_b = 0.4$\,\si{\nano\meter} and $r'_u = r_u = 0.9$\,\si{\nano\meter}, while for monomer--monomer binding/unbinding events the same cutoff lengths as in Model I are adopted. Distances are evaluated at intervals of $\Delta\tau = 0.5$\,\si{\nano\second}.

In both models, supramolecular aggregates are identified from the network of bonds between the core beads of distinct monomers. For the sake of clarity, assembly size distributions are reported using the following size clusters: monomer, dimer, 3--4, 5--8, 9--16, 17--32, 33--64, 65--128, and 129--256 monomers.

\subsubsection*{Supramolecular dynamics characterization and local supramolecular activity}
The molecular traffic, $T(t)$, is defined as the cumulative number of polymerization and depolymerization events per monomer up to time $t$~\cite{grant2011}. The time derivative, $\mathrm{d}T/\mathrm{d}t$, provides a direct measure of supramolecular activity. Polymerization and depolymerization probabilities as a function of the assembly size are calculated following the transition-matrix method described in Ref.~\cite{crippa2025nontrivial}; further details are provided in the SI.

Local quantities along the $z$ direction are evaluated using bins of width $0.02 L_{z}$. The local monomer and ion densities are obtained by averaging their spatial distributions over the final 20\,\si{\micro\second} of each trajectory. To characterize the coupling between local composition and supramolecular kinetics, binding and unbinding events, identified as described above, are assigned to the spatial bin in which they occur. The local activity is then quantified from the corresponding binding and unbinding event densities and related to the average monomer and ion densities in the same bin. 

\section*{Data availability}
Simulation input files, trajectories, and analysis scripts supporting the findings of this study are available in Zenodo (10.5281/zenodo.23340996).

\section*{Acknowledgements}
C.P. and A.T. acknowledge funding from the Italian Ministry of University and Research pursuant to Decree No. 23314 (11-12-2024)—BANDO FIS 2, SYSMAGRAD Project Grant No. FIS-2023-00936.

\section*{Competing interests statement}
The authors declare no competing interests.

\bibliography{bibliography}

\end{document}


\doublespacing

\maketitle

\section*{Polymerization and depolymerization probability analysis}

The size-dependent polymerization and depolymerization probabilities shown in Figs.~S1--S2 are computed from the simulation trajectories using the transition matrix method introduced in Ref.~\cite{crippa2025nontrivial}, which we briefly summarize here.

Supramolecular assemblies are identified from monomer core-to-core distances using the two-cutoff criterion described in the Materials and Methods section, with trajectory conformations sampled every $\Delta \tau = 0.5$\,\si{\nano\second}. At each sampling step, every monomer is classified according to the size $n$ of the assembly it belongs to, and its assembly size at the next step is recorded. The resulting transitions are accumulated into a transition matrix $\mathbf{T}(\Delta \tau)$, whose entry $T_{ij}$ counts the total number of observed transitions from an assembly of size $i$ to one of size $j$ over the interval $\Delta \tau$. By row-normalizing $\mathbf{T}(\Delta \tau)$ we obtain a transition probability matrix $\mathbf{P}(\Delta \tau)$, where $P_{ij}$ is the probability that a monomer in an assembly of size $i$ transitions to one of size $j$ in a single step. The transition probability matrix over a longer interval $N_{\mathrm{step}}\Delta\tau$ is then obtained via matrix multiplication:
\begin{equation}
  \mathbf{P}(N_{\mathrm{step}}\Delta\tau) = \bigl[\mathbf{P}(\Delta \tau)\bigr]^{N_{\mathrm{step}}}
\end{equation}
We use this relation to compute transition probabilities over $25$\,\si{\nano\second} ($N_{\mathrm{step}} = 50$ steps). The polymerization probability $P_\mathrm{pol}(n)$ and depolymerization probability $P_\mathrm{depol}(n)$ for assemblies of size $n$ are defined as:
\begin{align}
  P_\mathrm{pol}(n)   &= \sum_{j > n} P_{nj}(N_{\mathrm{step}}\Delta\tau), \\
  P_\mathrm{depol}(n) &= \sum_{j < n} P_{nj}(N_{\mathrm{step}}\Delta\tau).
\end{align}
Assembly sizes are coarse-grained into logarithmic bins for visualization, and error bars represent the standard deviation across independent trajectory blocks. %
\clearpage
\section*{Characterization of homogeneous simulations}

\begin{figure}[H]
  \centering
  \includegraphics[width=\linewidth]{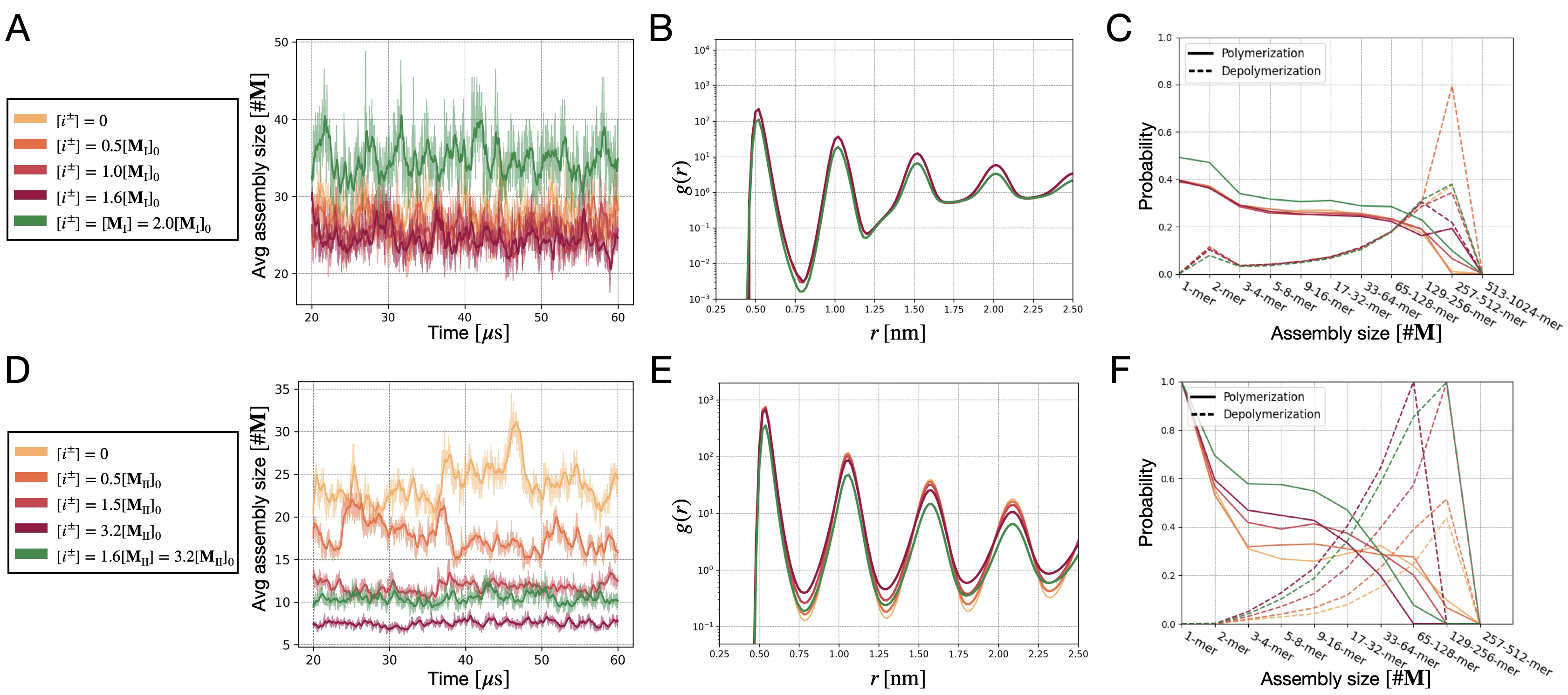}
  \caption{\textbf{Structural and kinetic characterization of homogeneous simulations.}
    For all homogeneous runs of Model~I (top panels: A,B,C) and Model~II (bottom panels: D,E,F):
    (A,D)~average assembly size $\langle n \rangle$ from 20 to 60\,\si{\micro\second}, with the dashed vertical line marking the start of the analysis window;
    (B,E)~radial distribution function $g(r)$ of the monomer core beads;
    (C,F)~size-dependent polymerization and depolymerization probabilities $P_\mathrm{pol}(n)$ and $P_\mathrm{depol}(n)$ over $\Delta\tau = 25$\,\si{\nano\second}, computed following Ref.~\cite{crippa2025nontrivial}. For the $[\mii] = 2.0\,[\mii]_0$ simulation only, the time axis of the average assembly size spans between $60$ and $100$\,\si{\micro\second}, because of its longer equilibration stage.
    Color coding as in Tables~1--2 of the main text.}
  \label{fig:s1}
\end{figure}
\clearpage
\section*{Characterization of all simulation runs}

\begin{figure}[H]
  \centering
  \includegraphics[width=\linewidth]{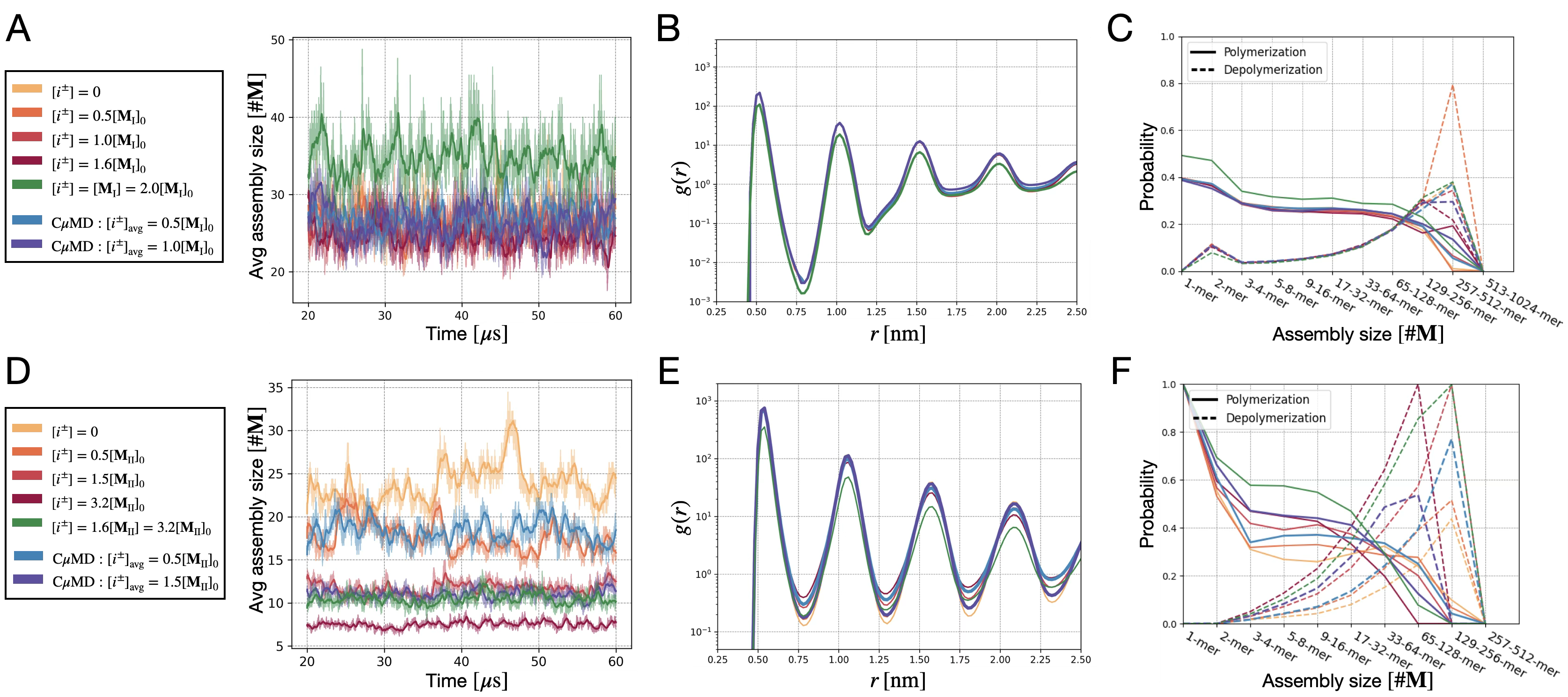}
  \caption{\textbf{Structural and kinetic characterization of homogeneous and C$\mu$MD gradient simulations.}
    Same observables as in Fig.~S1 — (A,D)~average assembly size, (B,E)~$g(r)$, (C,F)~$P_\mathrm{pol}(n)$ and $P_\mathrm{depol}(n)$ over $\Delta\tau = 25$\,\si{\nano\second} — shown for both homogeneous and C$\mu$MD runs of Model~I (top panels: A,B,C) and Model~II (bottom panels: D,E,F).  
    Color coding as in Tables~1--2 of the main text.}
  \label{fig:s2}
\end{figure}
\clearpage
\section*{Spatial profiles of supramolecular activity}

\begin{figure}[H]
  \centering
  \includegraphics[width=\linewidth]{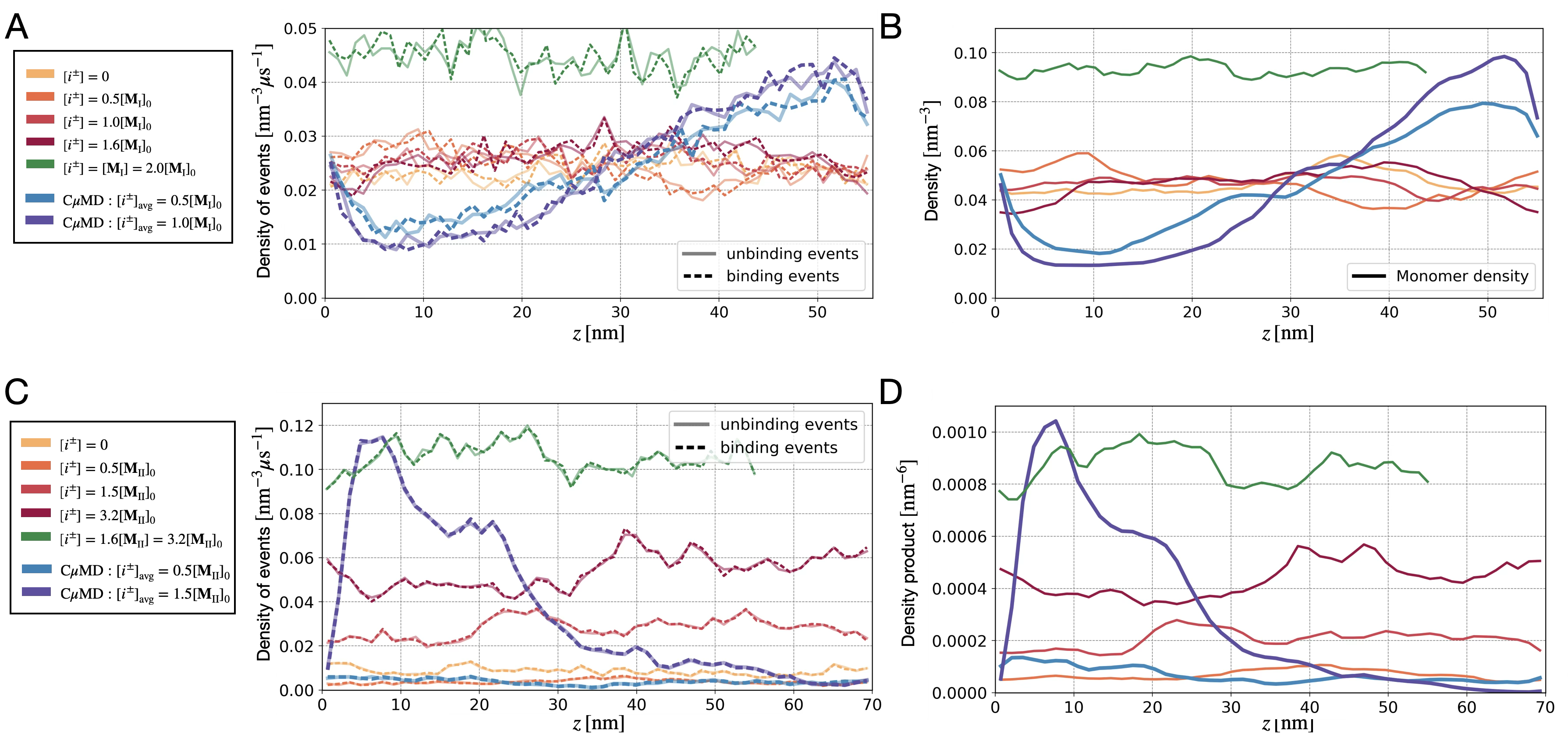}
  \caption{\textbf{Spatial profiles of binding and unbinding event densities along the gradient.}
    (A,~B)~Model~I: density of binding (solid) and unbinding (dashed) events along $z$ (A),
    and corresponding monomer density profile (B).
    (C,~D)~Model~II: density of binding and unbinding events along $z$ (C),
    and product of local monomer and ion linear densities (D).
    In Model~I the event profile tracks the local monomer density; in Model~II it tracks
    the product of local monomer and ion linear densities, reflecting the direct ionic coupling.
    Color coding as in Tables~1--2 of the main text.}
  \label{fig:s3}
\end{figure}

\bibliography{bibliography}